\documentclass{aa}  
\usepackage{graphicx}
\usepackage{newtxtext,newtxmath}
\usepackage[dvipsnames]{xcolor}

\def\ts     {\thinspace} 
\usepackage{amsmath}

\def\kms  {\ifmmode{{\rm \ts km\ts s}^{-1}}\else{\ts km\ts s$^{-1}$\ts}\fi}
\def\msol {\ifmmode{{\rm M}_{\odot}}\else{M$_{\odot}$\ts}\fi}
\def\lsun {\ifmmode{{\rm L}_{\odot}}\else{L$_{\odot}$\ts}\fi}
\def\cii  {\ifmmode{{\rm [C}{\rm \scriptstyle II}]}\else{[C\ts {\scriptsize II}]\ts}\fi}
\def\ci   {\ifmmode{{\rm C}{\rm \scriptstyle I}}\else{C\ts {\scriptsize I}\ts}\fi}
\def\m    {\ifmmode{\mu {\rm m}}\else{$\mu$m}\fi}
\def\ha   {\ifmmode{{\rm H}{\alpha}}\else{H$\alpha$\ts}\fi}
\def\hb   {\ifmmode{{\rm H}{\beta}}\else{H$\beta$\ts}\fi}
\def\ergcms   {\ifmmode{{\rm erg}\ts{\rm cm ^{-2}}\ts{\rm s ^{-1}}}\else{${\rm erg}\ts{\rm cm ^{-2}}\ts{\rm s ^{-1}}$\ts}\fi}
\usepackage{array}
\usepackage{amssymb}	
\def\ergcm2s{\ifmmode {\rm\,erg\,cm^{-2}\,s^{-1}}\else
                ${\rm\,ergs\,cm^{-2}\,s^{-1}}$\fi}
\def\ergsec{\ifmmode {\rm\,erg\,s^{-1}}\else
                ${\rm\,erg\,s^{-1}}$\fi}

\begin{document}

   \title{A resolved Ly$\alpha$ profile with doubly peaked emission at $z\sim7$}


   \author{C. Moya-Sierralta
          \inst{1}   
        \and
    J. González-López
    \inst{1,2,3}
        \and
    L. Infante
    \inst{1,2,3}
        \and
    L. F. Barrientos
    \inst{1}
        \and
    W. Hu
    \inst{4,5}
    \and
    S. Malhotra
    \inst{6}
        \and
    J. Rhoads
    \inst{6}
        \and
    J. Wang
    \inst{7,8}
        \and
    I. Wold
    \inst{6}
    \and
    Z. Zheng
    \inst{9}
    }
   \institute{Instituto de Astrof\'isica, Facultad de F\'isica, 
    Pontiﬁcia Universidad Cat\'olica de Chile, Santiago 7820436, Chile
              \\
              \email{cnmoya@uc.cl}
              \and
            Las Campanas Observatory, Carnegie Institution of Washington, Casilla 601, La Serena, Chile
            \and
           Instituto de Estudios Astrof\'isicos, Facultad de Ingenier\'ia y Ciencias,   Universidad Diego Portales, Av.  Ej\'ercito Libertador 441, Santiago, Chile [C\'odigo Postal 8370191]
            \and
            Department of Physics and Astronomy, Texas A\&M University, College Station, TX 77843-4242, USA
            \and
            George P. and Cynthia Woods Mitchell Institute for Fundamental Physics and Astronomy, Texas A\&M University, College Station, TX 77843-4242, USA
            \and
            Astrophysics Science Division, NASA Goddard Space Flight Center, 8800 Greenbelt Road, Greenbelt, Maryland, 20771, USA
            \and
            CAS Key Laboratory for Research in Galaxies and Cosmology, Department of Astronomy, University of Science and Technology of China, Hefei, Anhui 230026, People's Republic of China
            \and
            School of Astronomy and Space Science, University of Science and Technology of China, Hefei 230026, People 's Republic of China
            \and
            CAS Key Laboratory for Research in Galaxies and Cosmology, Shanghai Astronomical Observatory, Shanghai 200030, People's Republic of China
            }
   \date{}


  \abstract
   {The epoch of reionization is a landmark in structure formation and galaxy evolution. How it happened is still not clear, especially regarding which population of objects was responsible for contributing the bulk of ionizing photons toward this process. Doubly-peaked Lyman-Alpha profiles in this epoch are of particular interest since they hold information about the escape of ionizing radiation and the environment surrounding the source.}
   {We wish to understand the escape mechanisms of ionizing radiation in {\ifmmode{{\rm Ly}{\alpha}}\else{Ly$\alpha$\ts}\fi} emitters during this time and the origin of a doubly-peaked Lyman-alpha profile as well as estimating the size of a potential ionized bubble.}
   {Using radiative transfer models, we fit the line profile of a bright {\ifmmode{{\rm Ly}{\alpha}}\else{Ly$\alpha$\ts}\fi} emitter at $z\sim 6.9$ using various gas geometries. The line modeling reveals significant radiation escape from this system.}
   {While the studied source reveals significant escape ($f_{esc}$({\ifmmode{{\rm Ly}{\alpha}}\else{Ly$\alpha$\ts}\fi}) $\sim0.8$  as predicted by the best fitting radiative transfer model) and appears to inhabit an ionized bubble of radius $R_{b}\approx 0.8^{+0.5}_{-0.3}\,pMpc\left(\frac{t_{\rm age}}{10^{8}}\right)^{\frac{1}{3}}$. Radiative transfer modeling predicts the line to be completely redwards of the systemic redshift. We suggest the line morphology is produced by inflows, multiple components emitting Ly$\alpha$, or by an absorbing component in the red wing.}
   {We propose that CDFS-1's profile holds two red peaks produced by winds within the system. Its high $f_{esc}$({\ifmmode{{\rm Ly}{\alpha}}\else{Ly$\alpha$\ts}\fi}) and the low-velocity offset from the systemic redshift suggest that the source is an active ionizing agent. Future observations will reveal whether a peak is present bluewards of the systemic redshift or if multiple components produce the profile.}


    \maketitle
    
\section{Introduction}
The period following recombination, known as the Dark Ages, was characterized by a neutral intergalactic medium (IGM) and a lack of radiation sources. As structure formed, the first population of stars began to illuminate the cosmos, slowly ionizing the surrounding hydrogen. This process heated the pristine gas around the early galaxies, thus regulating the collapse and the formation of new stars. This marked the end of the Dark Ages and the beginning of the epoch of reionization \citep{loeb2001reionization,barkana2001beginning,furlanetto2005double,benson2006epoch,zaroubi2012epoch}.\\
The Epoch of Reionization (EoR) denotes a crucial phase during which most baryonic matter in the universe, particularly the IGM, became ionized, rendering it transparent to ionizing radiation. While approximate boundaries exist for the onset and conclusion of this epoch, debates persist regarding the specific mechanisms governing the reionization process \citep{Mason2018-du}.  Reionization is commonly described using three parameters that together determine $\Dot{n}_{ion}$, the influx of ionizing photons into the IGM. These are the ultraviolet luminosity density $\rho_{UV}$, which is closely tied to the number of galaxies hosting hot stars and is typically measured at rest wavelengths $\sim 1500$\AA; the ratio of ionizing photon production to UV continuum, $\xi_{ion}$; and the fraction of ionizing radiation escaping galaxies to reach the IGM ($f_{esc}$(LyC)) \citep{robertson2015cosmic}. Thus $\Dot{n}_{ion}$ can be written as:
\begin{equation}
    \centering
    \Dot{n}_{ion} \equiv \rho_{UV}\cdot \xi_{ion} \cdot f_{esc}(LyC)
\end{equation}
\\
The UV luminosity density  $\rho_{UV}$ can be directly determined from the Luminosity Function (LF) of galaxies observed at the epoch of reionization \citep{Bouwens2015,donnan2023evolution,harikane2023comprehensive,finkelstein2023ceers}. The conversion factor $\xi_{ion}$ from UV photons to ionizing photons can be derived from spectroscopic analysis of sources deemed to be similar to high redshift galaxies \citep{2017MNRAS.465.3637M,shivaei2018mosdef,Tang2019MNRAS.489.2572T}. More recently, advances facilitated by the James Webb Space Telescope (JWST) have enabled direct measurement of $\xi_{ion}$ from sources within the reionization era \citep{Atek2023-qf,matthee2023eiger,endsley2023jwst,prieto2023production,simmonds2023ionizing,simmonds2024low,simmonds2024ionisingpropertiesgalaxiesjades,saxena2024jades,pahl2024spectroscopic}. However, determining $f_{esc}$(LyC) poses challenges as residual neutral hydrogen in the IGM absorbs ionizing photons at redshifts $z>4$ \citep{vanzella2012detection,inoue2014updated}.

The behavior of $f_{esc}$(LyC) heavily influences models of the reionization history of the universe\citep{ishigaki2018full,finkelstein2019conditions}. Empirical relations have been proposed for $f_{esc}$(LyC) \citep{Izotov2016-rq,Izotov2018-xy,steidel2018keck,Matthee2021-jj,Pahl2021-kv,chisholm2022far,naidu2022synchrony,Pahl2023-hl,Pahl2024-ud}, and for $f_{esc}$({\ifmmode{{\rm Ly}{\alpha}}\else{Ly$\alpha$\ts}\fi}) \citep{Yang2016-yd,Sobral2019-te}, which correlates with $f_{esc}$(LyC) \citep{izotov2024ly}. Because the geometry of the medium is the main regulator of $f_{esc}$(LyC), Ly$\alpha$ profiles appear as one of the most promising proxies for the escape of ionizing radiation given that it is also mediated by the same effects \citep{Verhamme2006-dh}. 
The radiative transfer of {\ifmmode{{\rm Ly}{\alpha}}\else{Ly$\alpha$\ts}\fi} photons has been thoroughly studied in the literature \citep{1990ApJ...350..216N,Ahn2001,Ahn2002,Ahn2003,2006A&A...460..397V,Gronke2015-qp,li2022deciphering}, revealing that high escape fractions are associated with low neutral hydrogen column densities ($\log \rm{N}_{H_{I}}$), low dust optical depths ($\tau_{dust}$) and low neutral gas covering fractions \citep{Kakiichi2021-zv,steidel2018keck}. {\ifmmode{{\rm Ly}{\alpha}}\else{Ly$\alpha$\ts}\fi} profiles for such systems show undisturbed lines with non-zero flux at restframe {\ifmmode{{\rm Ly}{\alpha}}\else{Ly$\alpha$\ts}\fi} wavelength. In contrast, gas-rich and dusty systems hold asymmetric lines with significant offsets with respect to restframe {\ifmmode{{\rm Ly}{\alpha}}\else{Ly$\alpha$\ts}\fi}.

{\ifmmode{{\rm Ly}{\alpha}}\else{Ly$\alpha$\ts}\fi} profiles with flux bluewards of the systemic redshift are particularly interesting as they allow for measurements of the separation between peaks. Such systems usually appear as a doublet with zero flux between the blue and red wings, i.e., the so-called doubly-peaked {\ifmmode{{\rm Ly}{\alpha}}\else{Ly$\alpha$\ts}\fi} profiles, however in the case of highly efficient radiation escape the flux between the peaks does not necessarily reach zero \citep{naidu2022synchrony}.
At higher redshifts, it has been found that the Ly$\alpha$ line profile is insufficient to infer the fraction of escaping ionizing photons \citep{Pahl2024-ud,Choustikov2024-du}. However, one can still use it to select potential leakers, as emission near the line center is a necessary condition for high $f_{esc}$(LyC). In particular, the fraction of Ly$\alpha$ flux close to the systemic redshift ($f_{cen}$ and the Ly$\alpha$ escape fraction appear as promising proxies to be used at high-z \citep{choustikov2024great}.
To date, three sources at redshifts $z>6$ with a doubly-peaked profile have been reported in the literature \citep{hu2016ultraluminous,2018,matthee2018,meyer2021double}, with significant ($f_{esc}$(LyC)$>20\%$) inferred on all of them as measured from their peaks separation. The leading hypothesis as to why the blue wing of the line manages to reach the observer is an ionized bubble scenario in which the sources sit within an ionized hydrogen region with non-zero transmission at wavelengths shorter than $1215.6$\AA. Due to the Hubble flow, photons are redshifted as they move through the bubble, eventually being redshifted outside of the neutral hydrogen absorption range; thus, it is possible to infer lower limits of the bubble size by measuring the extension of the {\ifmmode{{\rm Ly}{\alpha}}\else{Ly$\alpha$\ts}\fi} blue wing \citep{Mason2020-lz}.\\

Resolving {\ifmmode{{\rm Ly}{\alpha}}\else{Ly$\alpha$\ts}\fi} profiles is still a promising technique to study the leakage of ionizing radiation at high redshift; however, technical limitations have made it difficult to carry out these observations. Optical and UV emission lines from EoR sources are observable with the James Webb Space Telescope (JWST). However, its instruments' spectral resolution cannot resolve the double peaks commonly observed in {\ifmmode{{\rm Ly}{\alpha}}\else{Ly$\alpha$\ts}\fi}\ profiles ($R\sim 2700$ or $\Delta v \sim 150$ km s$^{-1})$. On this end, ground-based instrumentation is currently the only way to conduct experiments to measure the escape of radiation as traced by {\ifmmode{{\rm Ly}{\alpha}}\else{Ly$\alpha$\ts}\fi} profiles. 

For these reasons, we have begun a spectroscopic campaign to follow up on the brightest confirmed sources from the LAGER narrow-band survey which aims to study reionization by selecting LAEs at $z\sim7$ to compute the Ly$\alpha$ luminosity function and the clustering of Ly$\alpha$ sources. Medium resolution spectroscopy with NIR coverage allows one to observe both {\ifmmode{{\rm Ly}{\alpha}}\else{Ly$\alpha$\ts}\fi} and nebular emission simultaneously, meaning that we can potentially resolve the {\ifmmode{{\rm Ly}{\alpha}}\else{Ly$\alpha$\ts}\fi} profile and also measure a systemic redshift to constraint the {\ifmmode{{\rm Ly}{\alpha}}\else{Ly$\alpha$\ts}\fi} emission line model.\\

We present new medium resolution data of a bright ($\log\frac{L_{\rm Ly}{\alpha}}{\ergsec}\approx 43.3$) LAE from the LAGER survey, namely CDFS-1, first reported in \cite{Hu2019-rg} and followed up by \cite{yang2019lyalpha}. We wish to understand the origin of CDFS-1’s doubly peaked Lyman-Alpha line and consider different scenarios that can give rise to it.

The paper goes as follows: Section 2 presents observations and data reduction. Section 3 presents the methodology and results. Section 4 is left for discussion, and section 5 will be used for conclusions. Throughout the paper, we use the cosmological parameters from the 2018 Planck results \citep{aghanim2020planck} 
$\mathrm{H}_{0}=67.4\,\kms\,{\mathrm{Mpc}}^{-1}$ and $\mathrm{\Omega}_{m}=0.315$

\section{Data}

\subsection{LAGER survey and sample}
The Lyman-Alpha Galaxies in the Epoch of Reionization survey (LAGER) is an ongoing project looking to detect $\sim 600$ LAEs at $z\sim7$ across 24 squared degrees. LAGER uses a custom narrow band filter centered on 964 nm mounted on the DECam at the Blanco 4m \citep{Zheng2017-jt,Zheng2019-wk}. Thanks to the combination of a large field of view and near-infrared sensitivity, LAGER has already selected hundreds of LAEs within the EoR and spectroscopically confirmed dozens of them, making it the largest LAE survey at this redshift range.

\begin{figure}
    \centering
    \includegraphics[width=\columnwidth]{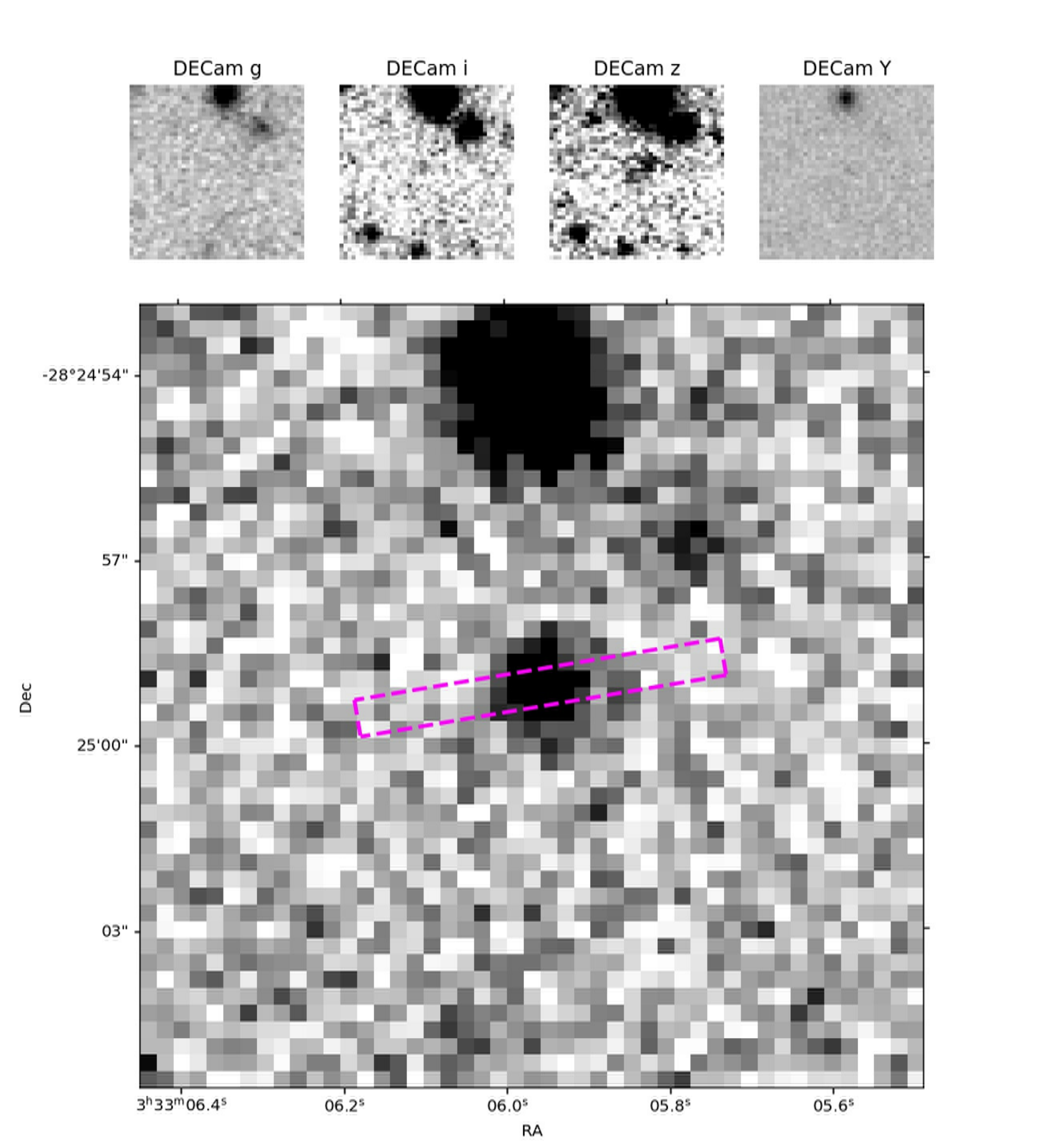}
    \caption{DECam g, i, z, NB  \& Y stamps centered around CDFS-1. The slit used for the presented observations is drawn on magenta on top of the narrowband stamp. The stamps are 12 arcsec side.}
    \label{fig:stamps}
\end{figure}

Being a narrowband survey, the selection is fairly complete for sources with $EW > 25$ \AA \citep{Wold2021-pp}. High {\ifmmode{{\rm Ly}{\alpha}}\else{Ly$\alpha$\ts}\fi} equivalent width has been linked with higher {\ifmmode{{\rm Ly}{\alpha}}\else{Ly$\alpha$\ts}\fi} escape fractions, meaning that LAGER sources are likely ionizing agents during the EoR \citep{steidel2018keck,Sobral2019-te,Pahl2021-kv,flury2022low}. Narrowband candidates are selected based on color criteria combined with veto bands; the best candidates are then subjected to spectroscopic follow-up. The spectroscopic confirmation process convolutes the selection functions, as the mask design is a function of candidates' quality and clustering. Lastly, to study the {\ifmmode{{\rm Ly}{\alpha}}\else{Ly$\alpha$\ts}\fi} profile, the brightest confirmed sources are considered for medium-resolution spectroscopic follow-up, as fainter sources would lead to low SNR spectra given the increased resolving power.

\subsection{Magellan/FIRE Spectroscopy}
CDFS-1 was observed for a total of five hours on source distributed across the nights of December 12th to 14th of 2020, with Magellan/FIRE. FIRE \citep{simcoe2013fire} is a medium-resolution echelle spectrograph mounted on the 6.5m Baade telescope at Las Campanas Observatory. The spectrograph provides simultaneous coverage from 0.8 to 2.5 $\mu m$. We used a 0.6 arcsec slit, which yields a resolution of $\sim 50$ km s$^{-1}$. Each of the twenty exposures is 900s long. The source is nodded along the slit in ABBA cycles followed by an AB cycle of a telluric standard. We define two positions, "A" and "B," offset by 2 arcsec; the sources are then observed on position A followed by B. This allows for better sky subtraction. A ThAr lamp is observed between pointings to minimize flexure effects. Each set of four (ABBA) science exposures plus two (AB) telluric exposures is referred to as a block.  After discarding non-useful data (e.g., pointing problems, poor conditions), we are left with 5 blocks totaling 5 hours of exposure time.

\subsection{Data Reduction}
Each block is reduced individually using the \textbf{FIREHOSE} IDL  library. We use the \textbf{fire\_addrectified} routine to produce rectified 2D spectra of each individual exposure as well as for tellurics. Since no continuum is expected for high-z LAEs, we use custom Python routines to stack individual exposures centered around the telluric's trace. We first create stacks for each unique block to minimize uncertainties from the separation of CDFS-1 and the telluric star in the image plane; each block stack is visually inspected to determine its position along the spatial direction. We then stack each block centered around the previously measured position. Lastly, the spectrum is extracted in a 1 arcsec aperture centered on the emission line.

The next step is to measure the line spread function (LSF) by comparing the width of skylines on our data and the skylines observed with higher resolution instruments (VLT/UVES, Program id:67.A-0278, P.I: López). We turn to the error array of our data, as we expect noise to be dominated by skylines in regions containing them. Thus, the width of the error spikes is similar to that of the skyline producing said noise. Since FIREHOSE creates sky-substracted data products, we use the error vector to measure the width of skylines.

We start by assuming that both the skylines and their trace on the error vector can be well modeled by a Gaussian function, the line width $\sigma_{lsf}$ is then  computed as a:\\

\begin{equation}
    \centering
    \sigma_{lsf} = \sqrt{\sigma_{data}^{2}-\sigma_{sky}^{2}}
    \label{eq:sigma}
\end{equation}

From equation \ref{eq:sigma}, we reach $\sigma_{LSF} = 60 \,\,\rm{km}\,\,\rm{s}^{-1}$, which is $20\%$ larger than the reported value for the 0.6 arcsec slit.\\
The point spread function (PSF) is measured by fitting a Gaussian function to the 2D telluric spectra collapsed in the spectral direction around the wavelength of interest (9634\AA). Consistent with the median seeing of the observing run, we measure a full width at half maximum (FWHM) of 0.6 arcsec.

\begin{figure*}
    \centering
    \includegraphics[width=\textwidth]{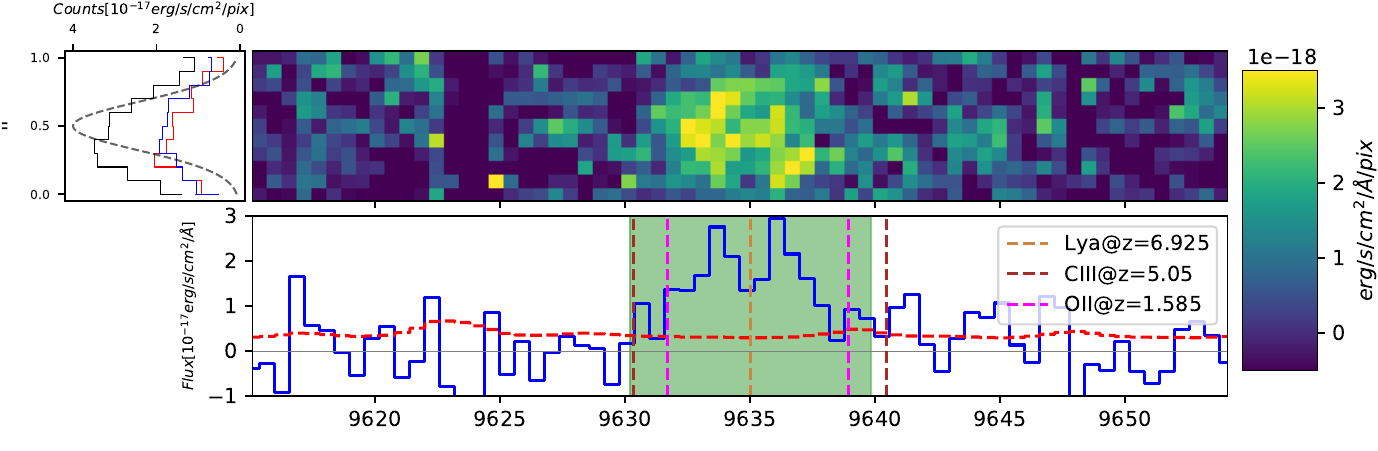}
    \caption{\textbf{Top} : Five hours combined 2D spectra centered around the {\ifmmode{{\rm Ly}{\alpha}}\else{Ly$\alpha$\ts}\fi} line. \textbf{Bottom:} Extracted 1D spectra. We sum all the flux within the columns. We also plot the expected positions of the [OII] and CIII] doublets, thus ruling out those lines as the origin of the observed double peak.  \textbf{Left:} Spatial profile of the emission line. We integrate the flux within the shaded region. We also plot only flux integrated bluewards (blue) and redwards(red) of 9635\AA.}
    \label{fig:spec2d}
\end{figure*}

\section{Methodology \& Results}
\subsection{Photometric properties}
We compute the {\ifmmode{{\rm Ly}{\alpha}}\else{Ly$\alpha$\ts}\fi} luminosity and the {\ifmmode{{\rm Ly}{\alpha}}\else{Ly$\alpha$\ts}\fi} equivalent width (EW) by combining DECam's z-band and narrowband photometry in tandem with the {\ifmmode{{\rm Ly}{\alpha}}\else{Ly$\alpha$\ts}\fi} redshift. The narrowband contains the line flux, while the z-band contains the line flux plus the stellar continuum. The {\ifmmode{{\rm Ly}{\alpha}}\else{Ly$\alpha$\ts}\fi} redshift tells us the filter transmission at the line wavelength. We create synthetic photometric measurements to estimate the uncertainties using the $1-\sigma$ limiting flux. We report the uncertainties as the 16th and 84th percentile ranges in table \ref{tab:CDFS1}.

\begin{table}
    \centering
    \begin{tabular}{c@{\vspace*{3pt}}c@{\vspace*{3pt}}}
        \hline
        Parameter & Measurement \\
        \hline
         $z_{{\ifmmode{{\rm Ly}{\alpha}}\else{Ly$\alpha$\ts}\fi}}$  & 6.9245\\
         $\log L$ & 43.3$^{+0.03}_{-0.04}$\\
         EW$_{{\ifmmode{{\rm Ly}{\alpha}}\else{Ly$\alpha$\ts}\fi}}$ & 79$^{+54}_{-23}$\\
         $f_{esc}$({\ifmmode{{\rm Ly}{\alpha}}\else{Ly$\alpha$\ts}\fi})$^{1}$ & 0.8$^{+0.09}_{-0.1}$\\
         \hline
         \end{tabular}
    \caption{Physical properties of CDFS-1. 1: {\ifmmode{{\rm Ly}{\alpha}}\else{Ly$\alpha$\ts}\fi} escape fraction from FLaREON radiative transfer modeling.}
    \label{tab:CDFS1}
\end{table}

\subsection{Line modelling}
As previously done in the literature \citep{matthee2018,hayes2021spectral}, we assume a systemic redshift such that {\ifmmode{{\rm Ly}{\alpha}}\else{Ly$\alpha$\ts}\fi} falls between the two peaks. The reasoning is that most doubly-peaked profiles indeed hold $z_{sys}$ between the peaks \citep{verhamme2018recovering}. Nevertheless, information on the systemic redshift is necessary to understand the mechanisms behind escaping radiation.

\begin{figure}
    \centering
    \includegraphics[width=\columnwidth]{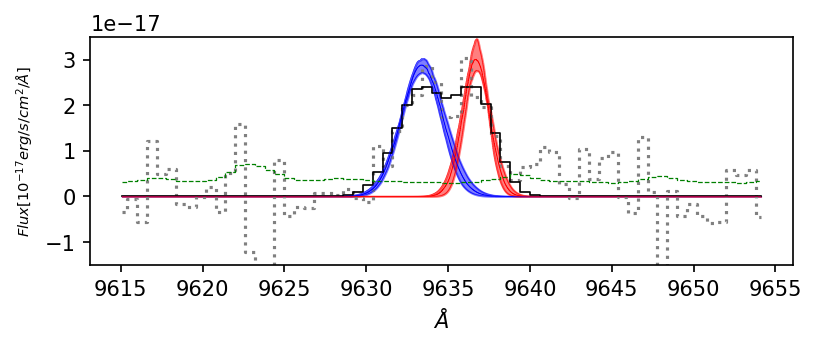}
    \caption{Result from fitting a double Gaussian to the data. The best fit shows a broader blue wing and a narrow red component. The shaded area represents the 16th and 84th percentile results from the MCMC routine. The solid black line is the sum of both components binned to match the data.}
    \label{fig:double_gauss}
\end{figure}

To find the emission line center, the total flux, and most importantly, the Blue-to-Red ratio, we start by fitting two Gaussian functions to our data as seen in figure \ref{fig:double_gauss}. We fit for the centroid, width, and amplitude of each Gaussianwith an MCMC approach using the emcee python library \citep{foreman2013emcee}. The MCMC fits two Gaussians with independent widths and centroids simultaneously. The Blue-to-Red ratio is then computed as the ratio of the areas of each Gaussian, and the systemic redshift is then computed from the average wavelengths of the peaks; results are shown in table \ref{tab:dg}.
\begin{table}
    \centering
    \begin{tabular}{l@{\vspace*{3pt}\hspace{10pt}}c@{\vspace*{3pt}\hspace{10pt}}c@{\vspace*{3pt}\hspace{10pt}}}
    \hline
    Paremeter & Blue & Red  \\
    \hline
    Amplitude[\ergcms \AA$^{-1}$]   & 2.466$^{+0.248}_{-0.195}$ & 3.811$^{+1.040}_{-0.696}$\\ 
    $\lambda$[\AA] & 9633.5$^{+0.151}_{-0.135}$ & 9636.58$^{+0.1}_{-0.1}$ \\
    $\sigma$[\AA] & 1.536$^{+0.561}_{-0.442}$&0.252$^{+0.188}_{-0.114}$ \\
    \hline
    \end{tabular}
    \caption{Results from fitting a double Gaussian model to the {\ifmmode{{\rm Ly}{\alpha}}\else{Ly$\alpha$\ts}\fi} emission line.}
    \label{tab:dg}
\end{table}

In order to distinguish between a double and a single peak, we compute the Bayesian Information Criteria (BIC) for a double and single Gaussian, respectively. We then measure $\Delta$BIC by subtracting the BIC of a double gaussian to the BIC of a single gaussian,
finding $\Delta$BIC$>10$, indicative of strong evidence towards a model \citep{raftery1995bayesian,szydlowski2015aic}. Thus there is strong evidence to choose a model with two Gaussians.
We then fit a radiative transfer model using FLaREON. FLaREON \citep{Gurung-Lopez2019-qa} is an open-source Python library that we used to construct models for the following geometries: Galactic Wind (GW) and Thin Shell (TS). The models are built from three free parameters for the GW and TS geometries. The FLaREON parameters are the medium expansion velocity (V$_{exp}$[km s$^{-1}$]), the neutral hydrogen column density of the shell ($\log\rm{N}_{H_{I}}$[cm$^{-2}$]) which is directly related to the number of scattering events a {\ifmmode{{\rm Ly}{\alpha}}\else{Ly$\alpha$\ts}\fi} photon suffers, and the dust optical depth ($\log\tau$) which regulates the number of absorption events. The models are computed from the \cite{Orsi2012-je} grid of radiative transfer simulations. The medium temperature is fixed at $T=10^{4} K$.

FLaREON’s thin shell model considers a point source surrounded by an isotropic shell of material at a fixed distance from the photon’s origin. Such configuration can be interpreted as radiation freely traveling before they interact with the ISM/CGM. Recently, more complex models that incorporate a multi phase medium have been suggested \citep{Kakiichi2021-zv,li2022deciphering}. Such models are able to more acutarely fit doubly-peaked profiles. We note however that the SNR presented in this paper limits our capability to distinguish between simple thin shell models and more sophisticated scenarios.

We fit FLaREON models to our data using the emcee library. Our fitting routine contains five free parameters: the three FLaREON parameters (V$_{exp}$, $\log \rm{N}_{H_{I}}$, and $\log\tau$), the systemic redshift, and a scale factor. The goodness of the fit is measured by the reduced $\chi^{2}$ computed from the residuals produced by subtracting each model to the data. Models are convolved with our estimated LSF before subtracting. FLaREON also yields $f_{esc}$(Ly$\alpha$) values for each fitted models. This is done by measuring the number of escaping photons divided by the total photon count in the intrinsic spectrum before radiative transfer.

\begin{figure}
    \centering
    \includegraphics[width=\columnwidth]{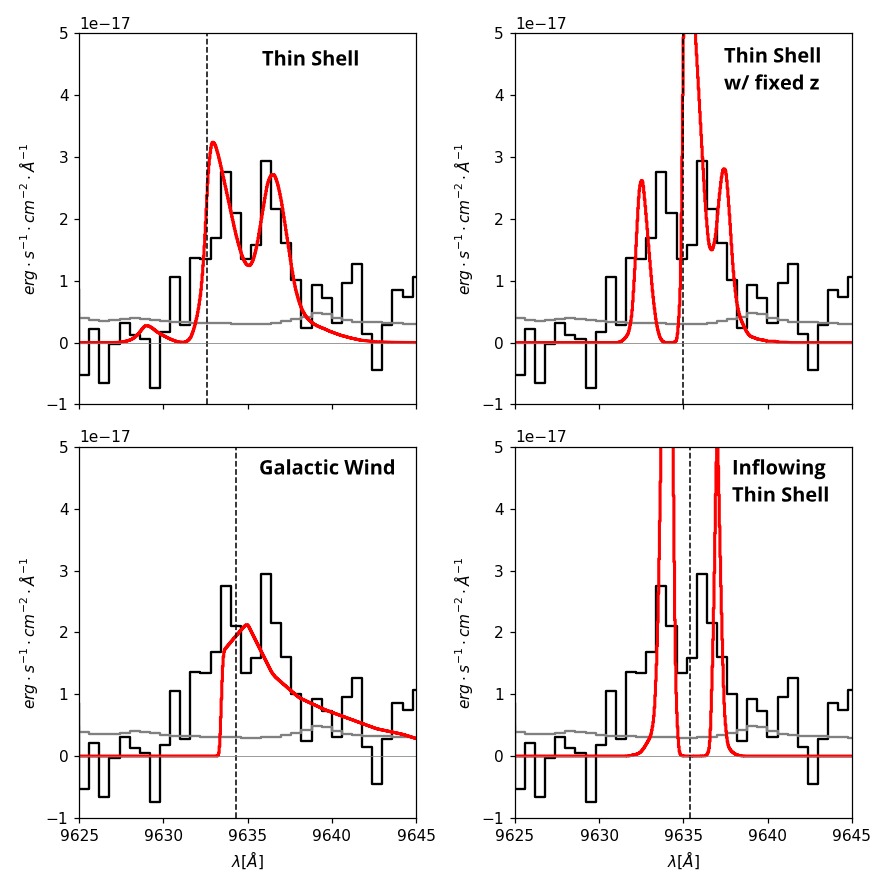}
    \caption{The four models we fit to the data. The best model for each geometry is plotted in red. Using a thin shell geometry with a free redshift yields the closest matching line profile. Models with a systemic redshift between the two peaks fail to match the narrow separation seen in CDFS-1. The galactic wind geometry fails to reproduce a doubly-peaked profile.}
    \label{fig:CDFS1_models}
\end{figure}

We fit four scenarios: A thin shell model with a free systemic redshift, a thin shell model with a fixed systemic redshift computed from the double Gaussian fitting, and a galactic wind model with a free redshift. By doing this, we wish to test if CDFS-1 is a bona fide double peak or if its profile arises due to other phenomena. We notice that the galactic wind solution does not present two distinct peaks. Thus, we will drop it from future discussions. The thin shell model with free systemic redshift offers satisfying solutions. However, when we fix the redshift to fall between the peaks, we obtain solutions with B/R ratios well below unity. We also use an inflow geometry to mirror the flux array concerning the systemic redshift. This inflow geometry allows us to fit negative velocities, i.e., infalling gas. For more details on the models, readers should refer to \citep{Gurung-Lopez2019-qa}.

\subsection{Comparison with other doubly-peaked LAEs}
 Evidence suggests CDFS-1 as the fourth doubly-peaked LAE at $z>6.5$ reported in the literature. In order to contextualize our results, we submit the other three sources to the same line modeling analysis. Moreover, we also run the fitting procedure on KRISS298 \citep{wofford2013lyalpha}. KRISS298 is a $z\sim 0.05$ LAE with an unusual $B/R>1$. By fitting FLaREON models to these four lines, we hope to validate the models and help us interpret the results from CDFS-1. The redshift is set free for all objects, adjusting the search range around their reported $z_{Ly\alpha}$.

The models manage to reproduce the doubly-peaked nature of COLA-1 and NEPLA-4. KISSR298 is not properly adjusted with an outflow geometry. Instead, an inflow geometry reproduces the observed profile.

The systemic redshift for the three sources is then fitted between the peaks, suggesting that FLaREON is capable of reproducing doubly-peaked {\ifmmode{{\rm Ly}{\alpha}}\else{Ly$\alpha$\ts}\fi} profiles. The geometric properties and the {\ifmmode{{\rm Ly}{\alpha}}\else{Ly$\alpha$\ts}\fi} escape fractions are reported in table \ref{tab:LAEs}. We note how all sources show similar $\log {\rm N_{H_{I}}}$ values, and for the high-z sources their $\log\tau$ are similar as well. 

Lastly, we note how the velocity offsets from the predicted systemic redshift and the width of the peaks in CDFS-1 are smaller than those of the other LAEs. This can be a signpost of efficient radiation escape, as we will discuss in the following sections.

\begin{figure}
    \centering
    \includegraphics[width=\columnwidth]{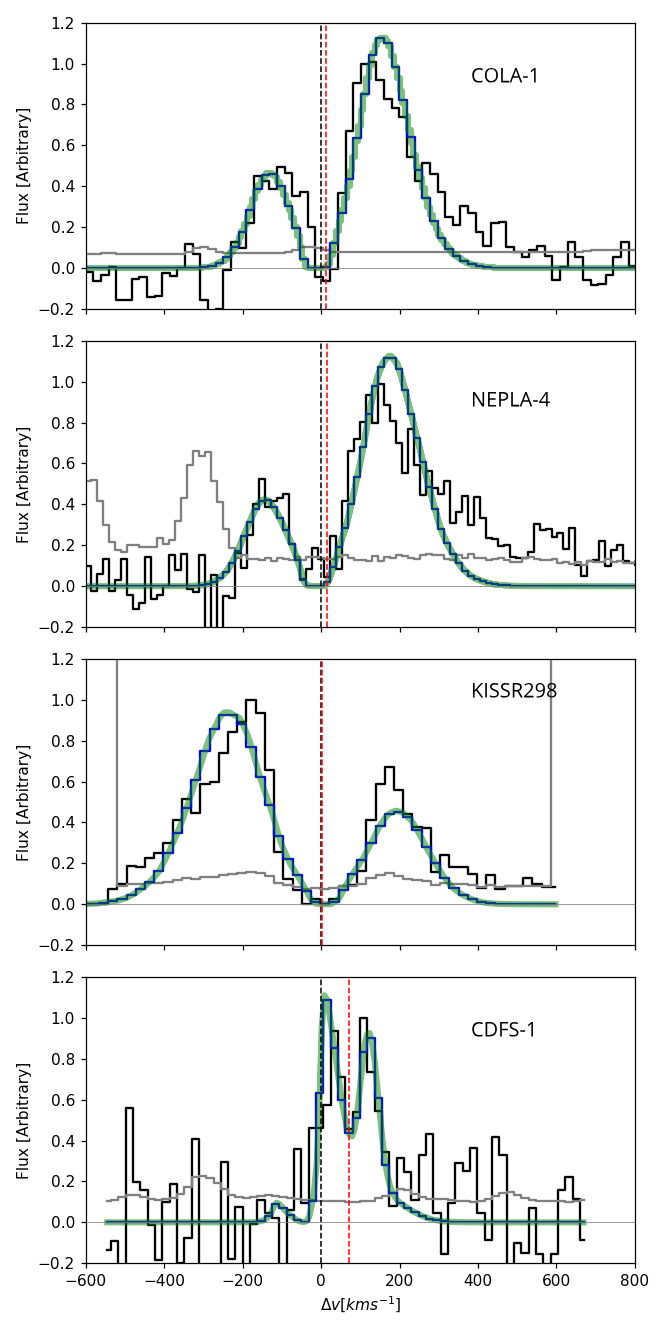}
    \caption{Radiative transfer models for {\ifmmode{{\rm Ly}{\alpha}}\else{Ly$\alpha$\ts}\fi} emitters in the literature. The systemic redshift is fitted between the peaks for every profile except for CDFS-1, suggesting that this source is indeed different from other LAEs. We also note that the peaks in CDFS-1 are visibly narrower than those of its counterparts. Black and red vertical dashed lines mark the zero and the best-fit expansion velocities, respectively.}
    \label{fig:rmodels}
\end{figure}
\begin{table}
    \centering
    \begin{tabular}{l@{\vspace*{3pt}\hspace{4pt}}c@{\vspace*{3pt}\hspace{4pt}}c@{\vspace*{3pt}\hspace{4pt}}c@{\vspace*{3pt}\hspace{4pt}}c@{\vspace*{3pt}\hspace{4pt}}}
         \hline
         Model & $z_{sys}$ & $\log {\rm N_{H_{I}}}$ & V$_{exp}$ & $\log\tau$  \\
         \hline
         Thin shell & 6.9237$^{+0.0002}_{-0.0002}$ & 18.3$^{+0.2}_{-0.5}$ & 70.1$^{+8.0}_{-5.5}$ & -2.47$^{+0.84}_{-0.52}$  \\
         Thin shell & 6.9257 (fixed) & 17.5$^{+0.02}_{-0.02}$ & 40.0$^{+12.5}_{-10.0}$ & -0.237$^{+0.031}_{-0.021}$ \\
         Galactic wind & 6.9251$^{+0.0001}_{-0.0001}$ & 19.75$^{+0.02}_{-0.02}$ & 450.0$^{+3.1}_{-4.7}$ & -0.103$^{+0.011}_{-0.011}$ \\
         Inflow & 6.9259$^{+0.0001}_{-0.0005}$ & 17.1$^{+0.1}_{-0.1}$ & 7.7$^{+3.6}_{-3.4}$ & -0.41$^{+0.1}_{-1.5}$  \\
         \hline
    \end{tabular}
    \caption{Results from fitting FLaREON models to our data. Errors are estimated from the spread in the posterior distribution. We find that the thin shell model with free redshift yields the best results.}
    \label{tab:FLRN}
\end{table}
\begin{table}
\begin{tabular}{l@{\vspace*{3pt}\hspace{4pt}}c@{\vspace*{3pt}\hspace{4pt}}c@{\vspace*{3pt}\hspace{4pt}}c@{\vspace*{3pt}\hspace{4pt}}c@{\vspace*{3pt}\hspace{4pt}}}
            \hline
           Source & $z_{sys}$ & $\log \rm{N}_{H_{I}}$ & V$_{exp}$ & $\log\tau$  \\
           \hline
           COLA-1 & 6.59$^{+0.003}_{-0.001}$ & 19.6$^{+0.03}_{-0.04}$  & 13.7$^{+2.0}_{-1.5}$ & -3$^{+0.001}_{-0.04}$ \\
           NEPLA-4 & 6.54$^{+0.004}_{-0.002}$ & 19.7$^{+0.06}_{-0.05}$  & 15.9$^{+4.0}_{-2.5}$ & -3$^{+0.001}_{-0.06}$\\
           KISSR298 & 0.048$^{+0.001}_{-0.003}$ & 19.7$^{+0.06}_{-0.03}$  & $-$2.28$^{+4.0}_{-3.9}$ & -0.76$^{+0.02}_{-0.05}$\\
           \hline
\end{tabular}
    \caption{Radiative transfer modeling solutions for two high (>6) redshift LAEs and a local LAE with a similar profile to CDFS-1. We use a thin shell model for all sources. In the case of KRISS298, we consider an inflowing thin shell, i.e., a thin shell flipped with respect to zero velocity.}
    \label{tab:LAEs}
\end{table}

\subsection{The bubble surrounding CDFS-1 }
We wish to extrapolate our findings to the broader population of LAEs at  $z\sim 6.9$. To do this, we first estimate the size of the ionized bubble surrounding CDFS-1. The equation for a bubble around an isolated ionizing source (ignoring recombination and the Hubble flow) is \citep{Mason2020-lz}:
\begin{equation}
    \centering
    R_{\rm ion} \approx \left(\frac{3f_{esc}(LyC)\Dot{N}_{\rm ion}t_{\rm age}}{4\pi\Bar{n}_{H}(z_{s})}\right)^{\frac{1}{3}}
    \label{eq:Rb}
\end{equation}
Where $t_{\rm age}$ is the timescale the source has been emitting ionizing photons, $\Dot{N}_{\rm ion}$ is the production rate of ionizing photons and $\Bar{n}_{H}(z_{s})= 1.88\cdot10^{-7}\cdot(1+z)^{3}{\rm cm^{-3}}$ is the mean cosmic hydrogen density at the source's redshift $z_{s}$. For the ionizing photon output $\Dot{\rm N}_{\rm ion}$ we turn to equation 5 from \cite{2018MNRAS.477.5406Y}:
\begin{equation}
    \centering
    {\rm L}_{{\ifmmode{{\rm Ly}{\alpha}}\else{Ly$\alpha$\ts}\fi}} = 0.68(1-f_{esc}(Lyc)) f_{esc}({\ifmmode{{\rm Ly}{\alpha}}\else{Ly$\alpha$\ts}\fi})\epsilon_{{\ifmmode{{\rm Ly}{\alpha}}\else{Ly$\alpha$\ts}\fi}}\Dot{\rm N}_{\rm ion}
    \label{eq:Yajima}
\end{equation}
Where $\epsilon_{{\ifmmode{{\rm Ly}{\alpha}}\else{Ly$\alpha$\ts}\fi}}$=10.2eV is the {\ifmmode{{\rm Ly}{\alpha}}\else{Ly$\alpha$\ts}\fi} transition energy. Combining \ref{eq:Rb}, and \ref{eq:Yajima} we get the following equation for a bubble of radius $R_{b}$:

\begin{equation}
    \centering
    {\rm R_{b}} = \frac{3.4\cdot 10^{-19}}{1+z}\cdot \left(\frac{{\rm L_{{\ifmmode{{\rm Ly}{\alpha}}\else{Ly$\alpha$\ts}\fi}}t_{\rm age}}f_{esc}(Lyc)}{(1-f_{esc}(Lyc))f_{esc}({\ifmmode{{\rm Ly}{\alpha}}\else{Ly$\alpha$\ts}\fi})}\right)^{\frac{1}{3}}
    \label{eq:bsize}
\end{equation}
\cite{izotov2024ly}  presented a tight empirical relation between $f_{esc}$({\ifmmode{{\rm Ly}{\alpha}}\else{Ly$\alpha$\ts}\fi}) and $f_{esc}$(LyC) by studying sources with Lyman continuum measurements in the form:
\begin{equation}
    \centering
    \log f_{esc}(Lyc) = 2.06473 \cdot \log f_{esc}({\ifmmode{{\rm Ly}{\alpha}}\else{Ly$\alpha$\ts}\fi})-0.0873
    \label{eq:izotov}
\end{equation}
Using our {\ifmmode{{\rm Ly}{\alpha}}\else{Ly$\alpha$\ts}\fi} luminosity, $f_{esc}$({\ifmmode{{\rm Ly}{\alpha}}\else{Ly$\alpha$\ts}\fi}), taking the escape fraction predicted by FLaREON, and $z_{{\ifmmode{{\rm Ly}{\alpha}}\else{Ly$\alpha$\ts}\fi}}$ to obtain $R_{b}\approx$0.8pMpc$(\frac{t_{\rm{age}}}{10^{8}})^{\frac{1}{3}}$.

Then, following a treatment similar to \citet{Malhotra2006}, we combined the bubble radius with the number density of {\ifmmode{{\rm Ly}{\alpha}}\else{Ly$\alpha$\ts}\fi}\ galaxies to derive a volume ionized fraction ($X_{\rm H_{II}}$).  We assume the same $f_{esc}$({\ifmmode{{\rm Ly}{\alpha}}\else{Ly$\alpha$\ts}\fi}) for all LAGER sources,  take the volume of a sphere with radius $r=R_{b}$, and integrate over the luminosity function to compute $X_{\rm H_{II}}$. 
We consider the LAE luminosity function from \cite{Wold2021-pp}. We integrate the corresponding volume expression for the radius given in \ref{eq:bsize} along the luminosity function, from $L_{*}$ to $\infty$ and assuming $t_{\rm age }=10^{8}$, in the same order of magnitude as the age of the universe at $z=6.9$.
\begin{equation}
    X_{\rm H_{II}} = \int_{L_{*}}^{\infty} \phi(L)R_{b}^{3}(L)dL
\end{equation}
We get a value of $X_{\rm H_{II}}\approx 0.25$, meaning that the ionized volume created by bright LAEs is insufficient to fully ionize the universe.

We note that this is an overestimation of the total ionized volume, as it assumes the same $f_{esc}$(Ly$\alpha$) for every source in the LAGER sample, as well as not taking into account overlap between bubbles.


\section{Discussion}
\cite{yang2019lyalpha} presented Magellan/FIRE spectra for CDFS-1 finding no emission lines other than {\ifmmode{{\rm Ly}{\alpha}}\else{Ly$\alpha$\ts}\fi}. The line is slightly/marginally resolved on their lower resolution data, hinting at a doubly-peaked emission line profile. The luminosity of CDFS-1 $L_{Ly\alpha} = 10^{43.3^{+0.03}_{-0.04}} ergs^{-1}$, coupled with its double-peaked profile, hints towards this source residing within an ionized region of the universe \citep{matthee2018,Mason2020-lz}. Aside from \cite{yang2019lyalpha}, spectroscopic confirmation of LAGER sources has been presented in \cite{Hu2017-vr}, \cite{Hu2021}, \& \cite{Harish2022-vp}.
While none of the spectra show a clear signal of a doubly-peaked profile, the spectral resolution of the observations is around $\Delta v \sim 200$ km s$^{-1}$. Figure 2 from \cite{yang2019lyalpha} shows CDFS-1 data in similar resolution with no hints of two peaks. This means that profiles such as the one seen in CDFS-1 might not be uncommon among the parent sample.

\subsection{Inference from radiative transfer modeling}

The results from various models fitted to CDFS-1 are presented in table \ref{tab:FLRN}, while the results of fitting LAEs in the literature are shown in table \ref{tab:LAEs}. Aside from the properties of the medium, FLaREON is able to compute $f_{esc}$(Lya), which is presented in table \ref{tab:CDFS1}.
CDFS-1 is different from the other sources in various aspects; it has an expansion velocity of $\sim 80$ km s$^{-1} $while the other sources hold small velocities $v_{exp}<20$ km s$^{-1}$. Its peaks are narrower, as seen in figure \ref{fig:rmodels}, and its predicted systemic redshift does not align with the gap between the peaks.
As the expansion velocity increases, the visibility of the blue peak decreases. This happens because photons can escape more easily in the red wing of the profile. However, large values of this parameter would lead to both red peaks merging onto a single wide wing \citep{Verhamme2006-dh}.

The narrower peaks seen in CDFS-1 can be attributed to the column density of neutral hydrogen. The column density predicted for CDFS-1 is more than an order of magnitude lower than the rest of the sources. A larger column density leads to a larger number of scattering events prior to escape, leading to photons distributing over a broader range in velocity space. Thus, $\log \rm{N}_{H_{I}}$ regulates both the width and the separation between peaks. 
We note that \cite{matthee2018} fitted radiative models to COLA-1, obtaining , $\log \rm{N}_{H_{I}} = 17.0$, as well as a temperature of $LogT = 4.2$ and an intrinsic line width of $\sigma_{int}=159 $ km s$^{-1}$. Including an intrinsic linewidth and a temperature alleviates the need for higher column densities to reproduce the observed line widths. Thus, we remind the reader that the results from radiative transfer are model dependent and conclusions should only come from comparing sources fitted with the same set of models.

Low-velocity solutions for CDFS-1 are unable to reproduce its B/R ratio, and while negative velocities are able to, they fail to match the width and positioning of the peaks (figure \ref{fig:CDFS1_models}). 
As measured by the reduced $\chi^{2}$, the best fit is a {\ifmmode{{\rm Ly}{\alpha}}\else{Ly$\alpha$\ts}\fi} profile with a double red peak. This scenario can reproduce the width, separation, and flux ratios between the peaks. The model predicts a gap at $\Delta v  \approx 70$ km s$^{-1}$, consistent with the expansion velocity of $v_{exp}=70.1$ km s$^{-1}$. This indicates that the expanding medium scatters photons near its {\ifmmode{{\rm Ly}{\alpha}}\else{Ly$\alpha$\ts}\fi} restframe wavelength while photons in the 'wings' can escape more easily. This interpretation does not present flux bluewards of the systemic redshift. However, this does not mean CDFS-1 lacks low-density channels for radiation escape. Its high $f_{esc}$({\ifmmode{{\rm Ly}{\alpha}}\else{Ly$\alpha$\ts}\fi}), predicted from either the {\ifmmode{{\rm Ly}{\alpha}}\else{Ly$\alpha$\ts}\fi} EW or from the FLaREON solution, coupled with its relatively low {\ifmmode{{\rm Ly}{\alpha}}\else{Ly$\alpha$\ts}\fi} velocity offset, suggest that photons are able to escape the system efficiently.

\subsection{Origin of the doubly-peaked profile}
A similar profile has been reported in \cite{Tang2024-zr}. The source identified as UDS-07665 shows Ly$\alpha$ emission close to the systemic redshift and fainter red wing separated by $\sim$100 km$s^{-1}$. The source also has an equivalent width similar to that of CDFS-1 EW(Ly$\alpha$) $\sim 100$ \AA. The line morphology is interpreted as a sign of efficient photon leakage given the high $f_{cen}$(Ly$\alpha$), similar to extreme sources such as the sunburst arc \citep{Rivera-Thorsen2017-pe} and Ion3 \citep{Vanzella2018-zc}. We note that this source possess a faint blue peak similar to the best fit FLaREON model, however the predicted flux is below the noise. We note however, that UDS-07665 has a central peak at systemic redshift while our best fitting model for CDFS-1 has a peak close to systemic but not at zero velocity. This could be produced by IGM absorption close to the Ly$\alpha$ rest-frame wavelength. If either a peak at systemic redshift or a clear peak at bluer wavelengths were detected, it would push or estimate of $f_{esc}$(Ly$\alpha$) higher as it would more strongly indicate the presence of escape channels within the galaxy.

We consider three scenarios: a) CDFS-1 has inflows, b) CDFS-1 contains at least two components emitting, and c)  CDFS-1 is a red wing with an absorption feature. 
The nature of the system is constrained by its unusually high B/R ratio. {\ifmmode{{\rm Ly}{\alpha}}\else{Ly$\alpha$\ts}\fi} double peak tends to show B/R ratios lower than one, especially at high redshifts where the IGM has a stronger attenuation towards the blue \citep{laursen2011intergalactic,hayes2021spectral}. Moreover, the IGM absorbs the blue peak even inside ionized bubbles \citep{Tang2024-zr}. Thus, a doubly-peaked Lya emitter requires a large bubble size or small separation to be observable. Given the stochasticity of the inter-galactic medium, we make no attempt to model it in this work and note that an LAE as bright as CDFS-1 is expected to host an ionized bubble. Combined with the fact that the profile is narrow with a small separation we justify the decision to ignore said component.

\cite{wofford2013lyalpha} presented a {\ifmmode{{\rm Ly}{\alpha}}\else{Ly$\alpha$\ts}\fi} double peak with a B/R ratio >1, which they interpret as infalling gas and is well fitted by our inflow model. Double peaks with B/R>1 have also been reported in \cite{Marques-Chaves2022-za}, \cite{Furtak2022-vz} and \cite{Mukherjee2023-zk}. However, their peak separations are considerably larger than that of CDFS-1. Similarly, \cite{Blaizot2023-zg} shows that profiles with B/R>1 do emerge in radiative transfer simulations, but once again, the separation between peaks is rarely below $\Delta v \approx 200$ km s$^{-1}$. Lastly, an inflow scenario seems unlikely given the unsatisfactory fit obtained with the inflow model.

The two peaks could arise from two distinct {\ifmmode{{\rm Ly}{\alpha}}\else{Ly$\alpha$\ts}\fi} emitting regions, which seems even more plausible considering that high redshift systems tend to be clumpier (for example \cite{fujimoto2024primordial}). While the two components seen on the 2D spectra are not co-spatial, they are still within the atmospheric FWHM, making their separation smaller than $d<5$ kpc. We use that distance in combination with the separation between peaks to estimate the mass of the system, yielding a value of $M_{dyn} = \frac{v^{2}\cdot d}{G} < 10^{10}M_{\odot}$. Were CDFS-1 to be two distinct emitting regions, one would like to distinguish between a merger and a clumpy system. Given the velocity value measured from the separation of the peaks in our double Gaussian fit ($\sim100$ km s$^{-1}$), a clumpy system appears more feasible. 
Restframe UV observations of high redshift LAEs reveal highly irregular morphologies. For example, the bright, narrow-band selected z>6 LAEs CR7 \& VR show multiple clumps in the UV while also showing spatial variations in their line profiles. \citep{Sobral2018-tq,Matthee2019-ic,Matthee2019-no,Matthee2020-vg}.
Martin et al. (in prep) present JWST/NIRCam imaging for 10 LAGER-selected LAEs. They find that 8/10 imaged sources are resolved into multiple components when observed at JWST's resolution.

\begin{figure}
    \centering
    \includegraphics[width=0.9\linewidth]{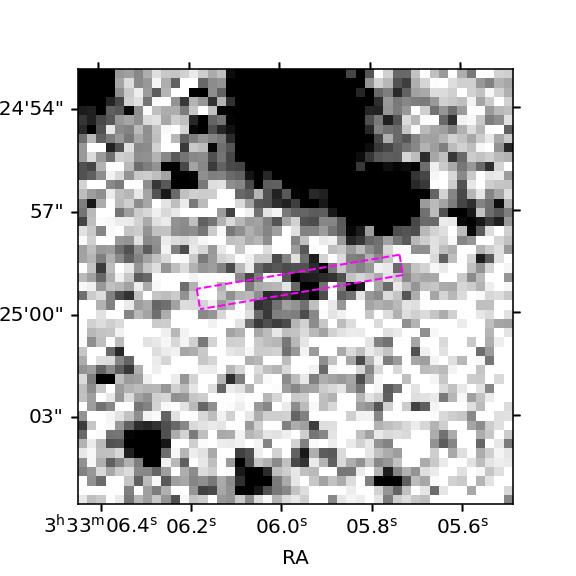}
    \caption{DECam z-band image of CDFS1. We overlay the slit as done in Figure 1. The morphology appears clumpy. While one of the components falls outside of the slit, the Ly$\alpha$ emitting region does not need to be co-spatial with the continuum emitting region.}
    \label{fig:z_decam}
\end{figure}

We note a hint of a second component in the z-band broadband data (see figure \ref{fig:z_decam}). However, just like the spectroscopic data, broadband imaging is limited by atmospheric seeing. Thus, we cannot rule out the possibility of CDFS-1 being an interacting system such as a merger. 
We propose the profile seen in CDFS-1 results from an expanding shell, which leaves an absorption feature on the {\ifmmode{{\rm Ly}{\alpha}}\else{Ly$\alpha$\ts}\fi} profile, resulting in a double ƒared peak. However, high-resolution imaging and a systemic redshift are required to rule out inflows and kinematic effects. This is likely a line-of-sight effect, as different sightlines might not hold reservoirs of neutral gas thus an absorption feature would not be produced.

\subsection{CDFS-1 as an ionizing agent}
Assuming the profile emerges from a single system, a doubly-peaked {\ifmmode{{\rm Ly}{\alpha}}\else{Ly$\alpha$\ts}\fi} profile with a systemic redshift between the peaks is a signpost of ionizing radiation escape. While this might or might not be true for CDFS-1, the redshift inferred from radiative transfer suggests a {\ifmmode{{\rm Ly}{\alpha}}\else{Ly$\alpha$\ts}\fi} offset of just $\sim 70$ km s$^{-1}$. Such a small offset is consistent with low-density channels that allow photons to escape undisturbed. This is further supported by the width of the peaks, which are visibly narrower than those of other LAEs. The high {\ifmmode{{\rm Ly}{\alpha}}\else{Ly$\alpha$\ts}\fi} equivalent width is also consistent with low-density channels, as {\ifmmode{{\rm Ly}{\alpha}}\else{Ly$\alpha$\ts}\fi} photons are more sensitive to surrounding neutral hydrogen than continuum photons\citep{Sobral2019-te,Pahl2021-kv}. 
Assuming the systemic redshift falls between the peaks, we can measure the minimum extension of the bubble surrounding CDFS-1. Following \cite{matthee2018} we measure a minimum radius $R_{min}\approx 0.2$ pMpc for a blue wing extending $\Delta v \approx 100$ km s$^{-1}$. This value is consistent with the radius computed in section 3.4 as long as $t_{\rm age}>10^{6}$ consistent with the expected age given the large {\ifmmode{{\rm Ly}{\alpha}}\else{Ly$\alpha$\ts}\fi} EW \citep{charlot1993lyman}. This means that CDFS-1 could produce a sufficiently large bubble to allow its blue peak to escape. The three previously known doubly-peaked LAEs at $z>6$ show comparable minimum radii of $R_{min}\sim 0.3$ pMpc consistent with the size of the bubbles powered by their {\ifmmode{{\rm Ly}{\alpha}}\else{Ly$\alpha$\ts}\fi} emission \citep{meyer2021double}. Thus, by fixing the systemic redshift between the peaks, one finds that all four of these sources are capable of ionizing their surrounding IGM, which in turn allows for the transmission of the blue peak.

\subsection{Hydrogen Neutral fraction at $z\sim 7$}
\begin{figure}
    \centering
    \includegraphics[width=\columnwidth]{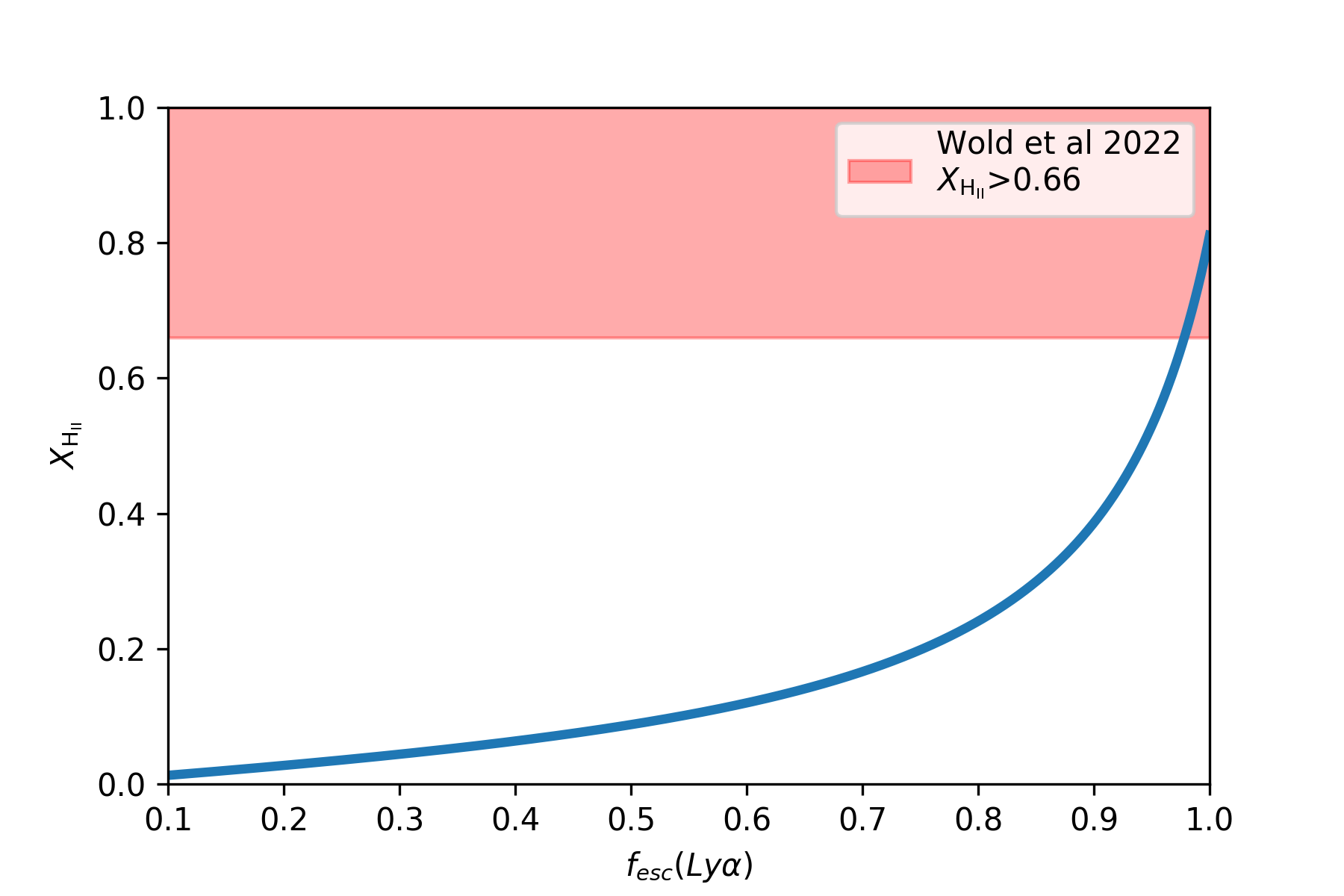}
    \caption{$X_{\rm H_{II}}$ as a function of $f_{esc}$({\ifmmode{{\rm Ly}{\alpha}}\else{Ly$\alpha$\ts}\fi}) at a redshift of $z\sim 6.9$. We integrate the volume of a bubble across the {\ifmmode{{\rm Ly}{\alpha}}\else{Ly$\alpha$\ts}\fi} luminosity function from $L_{*}$ up to $\infty$ assuming $t_{\rm age} = 10^{8}$. }
    \label{fig:fesc_vs_X}
\end{figure}

Assuming all bright LAEs have the same $f_{esc}$({\ifmmode{{\rm Ly}{\alpha}}\else{Ly$\alpha$\ts}\fi}) as CDFS-1, we find that they would ionize approximately a quarter of the IGM in the universe. Figure \ref{fig:fesc_vs_X} shows the relation between $X_{\rm H_{II}}$ and $f_{esc}$({\ifmmode{{\rm Ly}{\alpha}}\else{Ly$\alpha$\ts}\fi}) in this model. The figure shows that if bright ($L>L_{*}$) LAEs had {\ifmmode{{\rm Ly}{\alpha}}\else{Ly$\alpha$\ts}\fi} escape fractions close to 1, such as predicted by FLaREON for COLA-1, they'd be able to ionize most of the IGM by redshift $z=6.9$. 
The latest LAGER results measure a neutral hydrogen fraction  $X_{\rm H_{II}}>0.66$ at $z\approx 6.9$. Taking into consideration values from other probes, an estimate of  $X_{\rm H_{II}}>0.5$ is more conservative \citep{Wold2021-pp}.
If this volume were ionized exclusively by LAEs, they would require extreme conditions such as $f_{esc}$({\ifmmode{{\rm Ly}{\alpha}}\else{Ly$\alpha$\ts}\fi}) close to unity and timescales of  $t_{\rm age}\sim10^{8}$ years. Moreover, the {\ifmmode{{\rm Ly}{\alpha}}\else{Ly$\alpha$\ts}\fi} luminosity and escape fraction would need to remain constant during the burst period for bubbles to grow. $f_{esc}$({\ifmmode{{\rm Ly}{\alpha}}\else{Ly$\alpha$\ts}\fi}) was likely lower in earlier times, as the first generations of stars are required to carve escape paths for ionizing photons.

\section{Conclusions}

We have presented new spectra for CDFS-1, revealing a doubly-peaked profile. However, the profile differs from previously reported $z>6$ double peaks. In particular, we find a blue-to-red ratio higher than one and a small separation between the peaks of $\Delta v = 95.9 \pm 15.6$ km s$^{-1} $. This can be explained if the line morphology is not driven purely by radiative transfer. We propose three scenarios that would give rise to the observed emission: Inflating gas, kinematic effect, and absorption in the red wing of the profile.  Out of these scenarios, we choose the absorption on the red wing as the most plausible one. Our ground-based observations do not resolve the system into multiple components. Thus, there is no evidence supporting kinematic effects. For infalling gas, radiative transfer models cannot reproduce the observed profile. However, infalling gas can explain the profile of KISSR298 with a B/R ratio higher than one. 

The separation between peaks is considerably smaller than other double peaks observed at redshifts above six. Such a small gap is reproduced via an absorption on the red wing of the profile. The reported double-peak does not directly support the theory of an ionized bubble surrounding CDFS-1.
While the detection of flux bluewards of {\ifmmode{{\rm Ly}{\alpha}}\else{Ly$\alpha$\ts}\fi} signals the presence of an ionized region, the opposite is not true, as a bright LAE can power a surrounding bubble and show no hints of a double peak due to radiative transfer effects. The predicted {\ifmmode{{\rm Ly}{\alpha}}\else{Ly$\alpha$\ts}\fi} escape fractions coupled with the low {\ifmmode{{\rm Ly}{\alpha}}\else{Ly$\alpha$\ts}\fi} velocity offset from the predicted systemic redshift suggest an important contribution of ionizing photons towards the neutral IGM. If CDFS-1 were to be a bonafide doubly-peaked LAE, the minimum bubble size computed from the line profile (0.2pMpc) is consistent with the radius computed from its {\ifmmode{{\rm Ly}{\alpha}}\else{Ly$\alpha$\ts}\fi} luminosity and $f_{esc}$({\ifmmode{{\rm Ly}{\alpha}}\else{Ly$\alpha$\ts}\fi}), meaning the source can ionize a bubble large enough for the blue-peak to be observed.

Disentangling radiative transfer effects requires a systemic redshift to anchor the restframe {\ifmmode{{\rm Ly}{\alpha}}\else{Ly$\alpha$\ts}\fi} wavelength. This would immediately reveal whether CDFS-1 is a doubly-peaked LAE with a high B/R ratio and a narrow peak separation or a profile with two red peaks.
The systemic redshift could be obtained from ALMA observations of FIR lines or JWST spectroscopy of restframe optical emission lines. Moreover, such programs would also reveal information about the interstellar medium of ionizing sources during the EoR, allowing comparison with local analogs. As well as potentially revealing neighbouring systems that might also be pumping ionizing photons. 

To confirm or discard a multi-component scenario, IFU observations and NIR imaging are needed in order to resolve the line and continuum emission spatially. Once again, these observations can be achieved with both ALMA and JWST.

Lastly, we show how a population of bright LAEs with extreme {\ifmmode{{\rm Ly}{\alpha}}\else{Ly$\alpha$\ts}\fi} escape fractions ($f_{esc}\approx 1$) could ionize the universe by $z\sim 7$ given a long enough starburst period. $f_{esc}$({\ifmmode{{\rm Ly}{\alpha}}\else{Ly$\alpha$\ts}\fi}) is commonly observed to be lower than 1. Furthermore, our calculation assumes that this value and the {\ifmmode{{\rm Ly}{\alpha}}\else{Ly$\alpha$\ts}\fi} luminosity stay constant during the $10^{8}$ years timescale. These requirements suggest that reionization is not solely powered by such sources, and the contribution from other systems, such as fainter LAEs or AGNS, must be considered.

Future 30-meter class telescopes will be able to resolve the {\ifmmode{{\rm Ly}{\alpha}}\else{Ly$\alpha$\ts}\fi} profile of sources an order of magnitude fainter than what is currently possible. Studying those profiles will shed light on the radiative transfer effects and radiation escape fractions of such systems.

\section{Acknowledgements}
CMS acknowledges support from Beca ANID folio 21211528 and support from CATA. This paper includes data gathered with the 6.5-meter Magellan Telescopes located at Las Campanas Observatory, Chile. CMS and LFB acknowledge support from ANID BASAL project FB210003 and FONDECYT grant 1230231, and the China-Chile Joint Research Fund (CCJRF No. 1906). JXW acknowledges support from the science research grant from the China Manned Space Project with No. CMS-CSST-2021-A07. We want to thank Jorryt Matthee and Aida Wofford for sharing their data with us and aiding in the discussion. We also wish to thank the anonymous referee for constructive criticism. Based on observations collected at the European Organisation for Astronomical Research in the Southern Hemisphere under ESO program 67.A-0278.

\bibliographystyle{aa}
\bibliography{CDFS1}
\end{document}